# Multi-parameter mechanical and thermal sensing based on multi-mode planar photonic crystals

**Yongyao Chen, Haijun Liu, Zhijian Zhang, and Miao Yu**

*Department of Mechanical Engineering, University of Maryland, College Park, Maryland 20742, USA*

**Abstract**: This paper proposes a novel multifunctional sensing platform based on multimode planar photonic crystals (PPCs). We analytically and numerically demonstrate that the reflection spectrum of PPCs exhibits multiple high-Q resonant modes, and the fundamental and higher-order modes respond distinctively to external mechanical and thermal perturbations, rendering the PPCs superior capability for detection and discrimination of multiple parameters. We further demonstrate simultaneous pressure and temperature sensing with a PPC sensor. Other advantages of PPC sensors include on-chip integration and wafer-scale fabrications.

Simultaneous sensing of multiple parameters is highly desirable in many areas, including medical [1] and industrial applications [2]. Over the last two decades, advances in nanophotonics, novel materials, and micro/nano-fabrication techniques have led to significant progress in the development of miniature multifunctional optical sensors. These sensors often consist of many individual optical sensing elements integrated on a single chip. Each element is designed to measure a particular parameter, and therefore enables multi-parameter sensing on a single platform [3]. A major challenge faced by these sensors is usually the interferences among different measurands. For example, the measurement of strain, pressure, temperature, and ambient refractive index are often corrupted due to the cross-sensitivity to these parameters [3, 4]. Sensor isolation and compensation techniques are thus needed for these sensors to overcome the interference problem [3-5]. However, this will significantly increase the complexity as well as the cost of the sensing systems. In addition to on-chip optical sensors, fiber Bragg grating (FBG) sensors have also been reported for simultaneous measurements of strain, pressure, and temperature, owing to their capability of multiple parameter detection and discrimination [6, 7]. The principle is based on the fact that FBG sensors can support multiple resonance modes and each of these resonance modes responds differently with respect to different types of environmental perturbations [7]. However, these sensors often suffer from limited number of resonance modes due to the small refractive index variations in FBG structures, which limits the number of sensing parameters that can be detected [6, 7]. Moreover, the configuration and relatively long size of FBGs make them incompatible with cost-effective wafer-scale fabrication and integration. This hinders their applications as on-chip multi-parameter sensors and high spatial resolution sensing and imaging systems [8].

In this letter, we report a novel on-chip multi-parameter sensing platform based on planar photonic crystals (PPCs). PPCs are an important class of low-dimensional (one dimensional (1D) or two-dimensional (2D)) photonic crystal structures created on a planar substrate. These structures can be fabricated by using well-established micro- or nanofabrication techniques and are compatible with large-scale on-chip fabrication and integration [9]. The unique optical properties of PPCs include light-scattering resonance [10], strong optical confinement and high-Q resonance [11], which make them good candidates for the development of novel sensing modalities [12]. However, the multimode scattering resonance of PPCs and their potential for multi-parameter sensing have never been explored. Here, through analytical and numerical studies, we show that by taking the advantage of the synergetic effect of lattice and cavity resonances, the PPC structures can be designed to have high-Q, out-of-plane multimode reflection resonances that are highly sensitive to external perturbations. And more importantly, the fundamental and higher order resonant modes exhibit distinctive responses to different types of external stimuli. These properties render PPCs a novel on-chip multifunctional sensing platform for simultaneous measurement and discrimination of multiple parameters.

For proof-of-concept, we study a simple 1D polymer PPC structure deposited on a rigid substrate, as shown in Fig. 1(a). The periodic pattern can be transferred onto the surface of a polymer layer by using nano-imprinting lithography. The polymer film has a higher refractive index than the substrate, acting as a waveguide to confine the light. For a PPC structure with properly designed geometric parameters, when the PPC structure is illuminated with a broadband light source (normal incidence for instance), the reflection spectrum will exhibit multiple resonant modes at different wavelengths (see Fig. 1(b)). Moreover, as shown in Fig. 1(c), these resonances exhibit high Q-factors ($10^3 \sim 10^4$), which is particularly advantageous for the development of ultra-sensitive and high-resolution optical sensors [13].

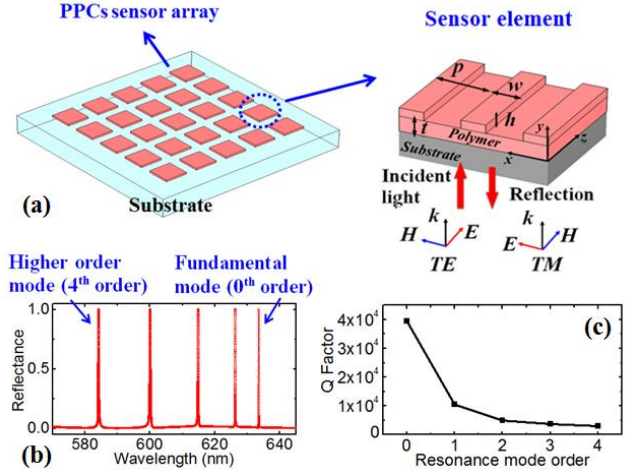

Fig. 1. (color online) Optical properties of PPCs. (a) Schematic of on-chip polymer based PPC sensors fabricated on a substrate (left). The inset (right) shows a single sensor element. The propagation of light is confined in the x-y plane. The environmental medium is air ($n_1$= 1). The PPC structure is made of polystyrene ($n_2$=1.59), and the substrate is fused silica ($n_3$= 1.45). The top periodic pattern has a thickness $h$ =100 nm, a grating width $w$ = 100 nm, and a periodicity $p$ = 400 nm. The bottom layer thickness of PPC structure is $t$ =2000 nm. TE or TM polarized light is incident from the bottom substrate. (b) Multimode resonance of PPCs for normally incident light with TE polarization. The reflection spectrum exhibits fundamental and higher order resonance modes. The simulation is based on rigorous coupled-wave analysis (RCWA). (c) The Q-factors corresponding to different resonance modes.

The field distributions of the fundamental mode and a higher order mode (4th mode) of the PPC structure are shown in Fig. 2(a). It can be observed that, under the resonance conditions, the incident light can be strongly coupled into the PPC structure and confined as guided modes in the polymer layer. From the analysis of field patterns, it is evident that the optical response relies on the synergetic effect of longitudinal and transverse resonances of the PPC structure, as schematically illustrated in Fig. 2(b). The top periodic structure of PPC acts as a coupler, which enables a momentum match between the external incident light and guided modes in the PPC structure, therefore, satisfying the longitudinal lattice resonance condition: $\beta = k \sin\theta + 2\pi m/p$. Here $\beta$ is the propagation constant of the guided mode, $k = 2\pi/\lambda$ is the wave-vector, $\theta$ is the incident angle, $p$ is the lattice period, and $m$ represents the diffraction orders. Note that $\beta$ is a longitudinal wave-vector along the PPC surface, and it can be quantified by the spatial periodicity of the mode field along x direction. As can be observed in Fig. 2(a), for normally incident light ($\theta = 0$), the fundamental and higher order modes have the same periodicity along the $x$ direction, satisfying the lattice resonance condition: $\beta = 2\pi/p$. On the other hand, the PPC layer also serves as a cavity along the transverse direction where the trapped light will reflect back and forth along the y direction, and thus the following transverse resonance condition is satisfied:

$$K_\perp d - \varphi_{21} - \varphi_{23} = N\pi \quad . \quad (1)$$

Here, $K_\perp$ is the transverse wave-vector that satisfies the relation: $K_\perp = [k^2 n_2^2 - (2\pi/p)^2]^{1/2}$. $d = h + t$ is the total thickness of the PPC structure with the pattern thickness of $h$ and the waveguide thickness of $t$. $n_1$, $n_2$ and $n_3$ represent refractive indices of the ambient medium (air), polymer PPCs, and substrate material, respectively. $N$ is a positive integer, indicating the resonance mode number. For instance, $N$=0 is the fundamental resonance mode, and $N$ >0 represents the higher order modes. $\varphi_{21}$ and $\varphi_{23}$ are phase shifts at the top and bottom interfaces of the PPC layer, and their values depending on the different light polarizations can be obtained as

$$\varphi_{2j} = \arctan\sqrt{\frac{(2\pi/p)^2 - (2\pi n_j/\lambda)^2}{(2\pi n_{2TE}/\lambda)^2 - (2\pi/p)^2}}, \; j\text{=1 or 3} \quad \text{TE} \quad (2a)$$

$$= \arctan\sqrt{\left(\frac{n_{2TM}}{n_1}\right)^4 \frac{(2\pi/p)^2 - (2\pi n_j/\lambda)^2}{(2\pi n_{2TM}/\lambda)^2 - (2\pi/p)^2}}. \quad \text{TM} \quad (2b)$$

Here, $n_{2TE}$ and $n_{2TM}$ represent the refractive indices of polymers for TE and TM polarizations, respectively (see polarization coordinates shown in Fig. 1(a)). For example, when a TE polarized light is incident on the PPC structure, the number of resonant modes and the corresponding resonance peak positions can be predicted by using Eqs. (1) and (2), as shown in Fig. 2(c), which are found to agree well with the numerical simulations by using rigorous coupled-wave analysis (RCWA) [14].

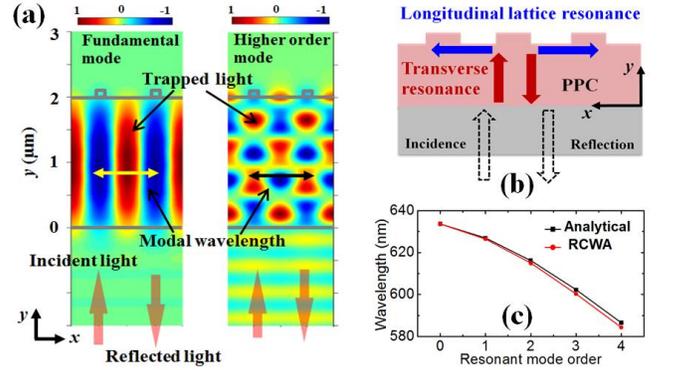

Fig.2. Resonance conditions of PPCs. (a) The electric field distributions of fundamental (left) and 4th order modes (right). The double-sided arrows indicate the spatial periodicity (modal wavelength) of the fundamental (yellow) and higher order modes (black). (b) Schematic of the synergistic effect of longitudinal and transverse resonant conditions. (c) The resonance wavelengths of PPCs corresponding to different resonant modes. The analytical results obtained by using Eqs. (1) and (2) are compared with the RCWA simulations.

We further investigate the influence of various environmental stimuli on the PPC resonance based on the above analytical model. As schematically shown in Fig. 3(a), environmental stimuli (e.g., pressure, strain, and temperature variations) can induce two different effects in the PPC structure. One is the structural deformations including the change in the PPC lattice periodicity and the thickness deformation of PPCs. Anther effect is the refractive index change due to the elasto-optic or thermal-optic responses of the ambient medium (air in our case), PPC polymer material, and substrate. Both effects can change the longitudinal and transverse resonance conditions in the PPCs, resulting in shifts of resonant peaks in the reflection spectrum. This is the key to the implementation of PPC based multi-parameter sensors. Specifically, the refractive index change due to elasto-optic or thermal-optic effects can be expressed as:

$$n_{jTE} = n_{jo} - \left[ C_{j1}\sigma_{jx} + C_{j2}(\sigma_{jy} + \sigma_{jz}) \right] + \alpha_j \Delta T \quad , \quad (3)$$

$$n_{jTM} = n_{jo} - \left[ C_{j1}\sigma_{jy} + C_{j2}(\sigma_{jx} + \sigma_{jz}) \right] + \alpha_j \Delta T \quad , \quad (4)$$

where the subscript $j = 1,2,3$ corresponds to the ambient medium (air), polymer PPCs, and substrate material (fused silica), respectively, $n_{jo}$ is the corresponding refractive index without applying any stimuli, $\sigma_{jx,y,z}$ represents external forces or thermal induced stress components in the corresponding material, $C_{j1}$ and $C_{j2}$ are elasto-optical constants of the corresponding material [15], $\alpha_j$ is the thermal-optic coefficient of the corresponding material, and $\Delta T$ is the temperature change of the environment. On the other hand, the longitudinal and transverse structural deformations of PPCs can be characterized as the following:

$$p = (1+\varepsilon_x)p_o, \quad (5)$$

$$d = (1+\varepsilon_y)d_o. \quad (6)$$

Here, $p_o$ and $d_o$ represent the lattice periodicity and PPC thickness without applying external stimuli, respectively, $\varepsilon_x$ and $\varepsilon_y$ denote the induced strains along the longitudinal and transverse directions, respectively. Substituting Eqs. (3) to (6) into the analytical model (Eqs. (1) and (2)), the mechanical and thermal effects on the optical resonance of PPC structures can be analyzed. Note that, due to the small elasto-optic and thermal-optic coefficients of the ambient medium (air) and the rigid substrate (fused silica) relative to the PPC polymer material [16, 17], the related refractive index variations in the air and the substrate can be neglected. Therefore, for simplicity, we only consider the refractive index change in the polymer PPC layer in the following analytical and numerical analyses.

In Figs. 3(b)-3(d), we show how various structural deformations and the induced refractive index change will influence the resonance shift of PPCs. It can be observed that for both the fundamental and 4th order modes, the resonant wavelengths shift linearly with respect to the polymer refractive index change (Fig. 3(b)) as well as structural deformations (e.g., periodicity change in Fig. 3(c) and PPC thickness change in Fig. 3(d). These results indicate that the optical response of PPCs can be exploited for various sensing applications, such as strain and stress monitoring, pressure sensing, and temperature detection. More importantly, the fundamental and higher order resonance modes of PPCs are found to respond differently with respect to the same external stimuli induced structural deformations or refractive index change. This is the key to the implementation of multimode PPCs as a novel sensing platform for simultaneous measurement and discrimination of multiple parameters. For example, for the simulated PPC structure shown in Fig. 1(a), since there are 5 resonance peaks in the wavelength range of interest, ideally, up to 5 different parameters can be measured simultaneously, which can be pressure, strain and stress fields, temperature, ambient refractive index change, etc. It should be noted that the similar sensing mechanism is also applicable to 2D PPC based devices.

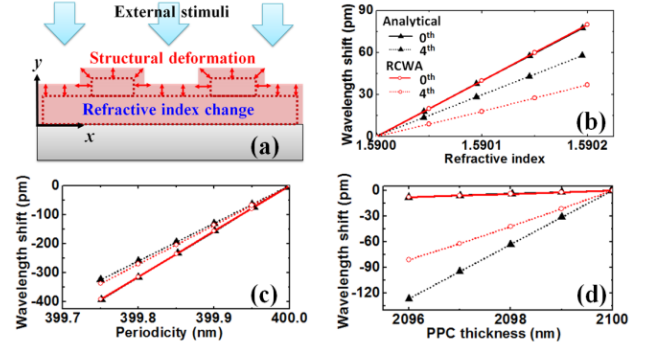

Fig.3. Influence of environmental perturbations on the PPC resonance. (a) Schematic of the external stimuli (e.g., pressure, strain and temperature) induced structure deformations as well as refractive index change in PPC structures. (b) Resonance shift with respect to the refractive index change of the polymer PPC structure. (c) PPCs (longitudinal deformation) resonance shift due to lattice variations. (d) PPC resonance shift due to thickness changes (transverse deformation). The analytical results are based on Eqs. (1) and (2), which are compared with the RCWA simulations.

As a practical example, we demonstrate the feasibility of simultaneous pressure and temperature sensing with a single PPC element, which may find important applications in the fields of medical diagnosis and industrial monitoring systems. For the device shown in Fig. 1(a), under external pressure (0 MPa to 2 MPa) and temperature (20 °C to 30 °C), the stress and strain components in the PPC structure are obtained by using the Finite Element Method. Substituting the obtained stress and strain components into Eqs. (3) to (6), the structural deformations as well as the induced refractive index change can be determined, which are further used in the RCWA simulations to analyze the PPC resonance shifts due to the mechanical and thermal stimuli. In Fig. 4, the resonant wavelength shifts with respect to the pressure and temperature (both induce deformation and refractive index changes) are obtained. For proof-of-concept, only the fundamental and the 4th order modes are studied. As shown in Fig. 4(a), the fundamental mode and the 4th order mode respond differently with respect to the applied pressure; the resonant wavelength of the fundamental mode exhibits a red shift (Fig. 4(b)), while the resonance wavelength of the 4th order mode has a blue shift (Fig. 4(c)). The pressure sensitivities based on the fundamental mode and the 4th order mode can be determined to be $K_{P0}$ = 16 pm/MPa and $K_{P4}$ = –80 pm/MPa, respectively, which are distinctively different. Further, we investigate the temperature response of the same PPC structure. For both the fundamental mode and the 4th order mode, a blue shift in the resonance wavelengths is observed as the temperature increases (see Figs. 4(d)-4(f)). However, although both the fundamental and higher order modes exhibit similar linear temperature response, their temperature sensitivities are quite different. The temperature sensitivity of the fundamental mode ($K_{T0}$ =–45 pm/°C) is much higher than that of the 4th order mode ($K_{T4}$ =–18 pm/°C).

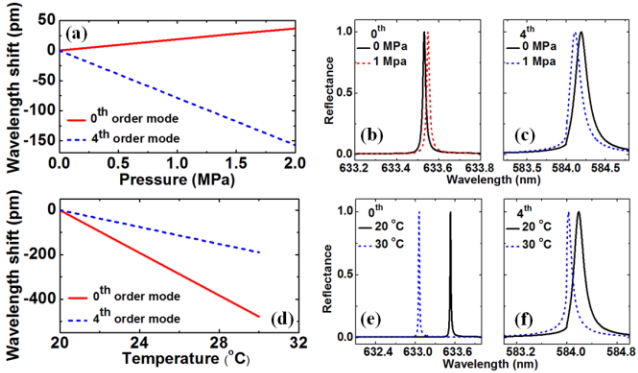

Fig.4. The optical resonance of PPCs subjected to pressure and temperature variations. (a) The resonance shift of the fundamental and higher order modes with respect to the applied pressure. The shift of the reflection spectrum under external pressure field for the fundamental mode (b) and the higher order mode (c). (d) The resonance shift versus temperature for the fundamental and higher order modes. The reflection spectrum shift of the fundamental mode (e) and the higher order mode (f) under the environmental temperature change.

Since both the pressure and temperature responses of the PPC sensor are linear, we can obtain the following measurand interaction equation:

$$\begin{pmatrix} \Delta T \\ \Delta P \end{pmatrix} = K_{TP}^{-1} \begin{pmatrix} \Delta \lambda_0 \\ \Delta \lambda_4 \end{pmatrix} = \begin{pmatrix} K_{P0} & K_{P4} \\ K_{T0} & K_{T4} \end{pmatrix}^{-1} \begin{pmatrix} \Delta \lambda_0 \\ \Delta \lambda_4 \end{pmatrix}$$
$$= \frac{1}{K_{T0} K_{P4} - K_{P0} K_{T4}} \begin{pmatrix} K_{P4} & -K_{P0} \\ -K_{T4} & K_{T0} \end{pmatrix} \begin{pmatrix} \Delta \lambda_0 \\ \Delta \lambda_4 \end{pmatrix}, \quad (7)$$

where $\Delta P$ and $\Delta T$ are pressure and temperature variations, $\Delta \lambda_0$ and $\Delta \lambda_4$ represent the corresponding resonance shifts of the fundamental and 4th order modes, respectively, and $K_{TP}$ is the sensitivity matrix of the PPC sensor. Therefore, once the PPC sensor is calibrated, by using the measured resonance shifts, simultaneous pressure and temperature detection and discrimination can be achieved.

In general, the detection of more parameters, such as three dimensional strain fields and the refractive index of surrounding medium, can also be realized by using the multiple resonant modes of PPCs. It is worth noting that the accuracy of multi-parameter sensing largely depends on the determinant of the PPC sensitivity matrix (e.g., $K_{T0} K_{P4} - K_{P0} K_{T4}$ for temperature and pressure sensing). To reduce measurement errors, the sensitivities in the matrix should be distinctively different [7]. Therefore, to successfully develop PPC multi-parameter sensors, not only the identification of multiple resonant peaks but also the distinctive sensitivities of the resonant modes are important. As discussed previously, the optical sensitivities of the fundamental and higher order modes of PPCs to structural deformations and refractive index change are considerably different, which makes PPCs a promising platform for the multi-parameter sensing.

In summary, we studied the optical resonance of multimode planar photonic crystals and demonstrated their potentials as a novel on-chip multifunctional sensing platform. It should be noted that the performance of PPC sensors, such as the Q factor of the resonant modes and the sensor sensitivity to environmental changes can be improved through optimization of geometric designs and material selections. Furthermore, even though the PPC sensor was designed to work in the optical region, the same working mechanism is also applicable in other spectrum regions, such as infrared, terahertz, and microwave regimes, where multi-parameter detection and discrimination are critically needed for many applications.